\documentclass{article}
\usepackage[preprint]{spconf}
\usepackage{amsmath,graphicx}
\usepackage{color}
\usepackage{booktabs}
\usepackage{multirow}
\usepackage{subcaption}
\usepackage{nicefrac}
\usepackage{tabularx}

\title{Severity classification in cases of Collagen VI-related myopathy with Convolutional Neural Networks and handcrafted texture features}
\name{\begin{tabular}{@{}c@{}}
        Rafael Rodrigues$^{1}$\thanks{Research  funded by FCT/MCTES, under the project UIDB/EEA/50008/2020 and the project CENTRO-01-0145-FEDER-000019 - C4 Cloud Computing Competence Centre. Author R. Rodrigues would also like to thank Funda\c{c}\~ao para a Ci\^encia e a Tecnologia (FCT/MCTES) for funding this research, under the doctoral grant SFRH/BD/130858/2017.}\quad 
        Susana Quijano-Roy$^{2}$\quad
        Robert-Yves Carlier$^{2}$ \quad 
        Antonio M. G. Pinheiro$^{1}$
\end{tabular}}

\address{$^{1}$ Instituto de Telecomunica\c{c}\~oes and Universidade da Beira Interior, Covilhã, Portugal \\
          $^{2}$ APHP - Raymond Poincar\' e University Hospital, Garches, France}

\toappear{\copyright \mbox{} 2022 IEEE}
         
\begin{document}
%
\maketitle
\begin{abstract}
Magnetic Resonance Imaging (MRI) is a non-invasive tool for the clinical assessment of low-prevalence neuromuscular disorders. Automated diagnosis methods might reduce the need for biopsies and provide valuable information on disease follow-up. In this paper, three methods are proposed to classify target muscles in Collagen VI-related myopathy cases, based on their degree of involvement, notably a Convolutional Neural Network, a Fully Connected Network to classify texture features, and a hybrid method combining the two feature sets. The proposed methods were evaluated on \mbox{axial} T1-weighted Turbo Spin-Echo MRI from 26 subjects, including Ullrich Congenital Muscular Dystrophy and Bethlem Myopathy patients at different evolution stages. 
The hybrid model achieved the best cross-validation results, with a global accuracy of 93.8\%, and F-scores of 0.99, 0.82, and 0.95, for healthy, mild and moderate/severe cases, respectively.
\end{abstract}
\section{Introduction}
Neuromuscular diseases (NMD) comprise a wide range of individually rare disorders, with several different causes and phenotypes \cite{ten2016muscle, quijano2019neuroimaging}. Ullrich Congenital Muscular Dystrophy (UCMD) is an inherited early-onset disorder, caused by mutations in the Collagen VI (COL6) genes, and is recognized as the most severe form of COL6-related myopathy. Bethlem Myopathy corresponds to a milder form, which typically has a later onset \cite{mercuri2005muscle}. The observed symptoms include generalized muscle weakness and hypotonia, joint contractures, distal joint hyperlaxity and scoliosis. Motor development is often delayed, and walking ability might be lost or never acquired, in the most severe forms. Also, in more severe cases, patients may end up developing spinal deformities and life-threatening respiratory insufficiency \cite{quijano2012whole, bazaga2019convolutional}.

The muscle involvement pattern is similar across the COL6 myopathy spectrum, albeit with varying degrees of severity \cite{ten2016muscle, mercuri2005muscle}. The appearance of a striped pattern, with alternating bands of hypointensity (i.e., preserved muscle) and hyperintensity (i.e., fat/connective tissue), is common in T1-weighted (T1\emph{w}) Magnetic Resonance Imaging (MRI) \cite{quijano2012whole}. In the thigh, muscle involvement is typically diffuse, with some common patterns in T1\emph{w} MRI, such as a hyperintensity rim appearing in the \textit{Vastus lateralis} and central hyperintensity ("target") in the \textit{Rectus femoris} \cite{mercuri2005muscle}.

\begin{figure}[t!]
\centering
\includegraphics[width=0.95\linewidth]{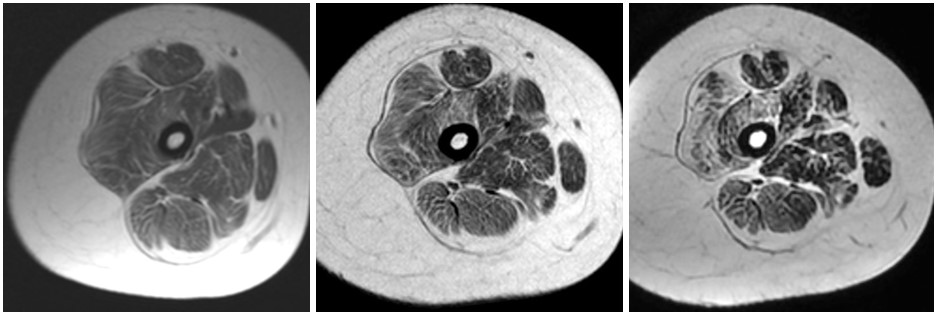}
\caption{Progression of COL6 myopathy in T1\emph{w} MRI exams of an ambulatory young female affected with UCMD, at ages 5, 12 and 19 (from left to right).} \label{fig:progression}
\end{figure}

MRI provides a reliable and non-invasive clinical outcome measure for the diagnosis and monitoring of NMD \cite{quijano2019neuroimaging, mercuri2005muscle}. In specialized centers, it may be possible to assess the course of the myopathy using muscle MRI (Fig. \ref{fig:progression}), helped by standardized scoring forms of signal and volume changes \cite{quijano2019neuroimaging}. Texture analysis of MRI scans has shown great potential in the development of computer-aided diagnosis (CAD) methods for NMD that might ultimately help reduce the need for biopsies, which are currently widely used in diagnosis and clinical follow-up of patients. However, developing CAD methods for these diseases is still a very challenging field, particularly due to their individual rareness \cite{quijano2019neuroimaging, de2015application}. In this paper, we propose the use of Convolutional Neural Networks (CNN) to classify different stages of muscle involvement, on T1\emph{w} Turbo Spin-Echo (TSE) MRI scans of UCMD and Bethlem Myopathy patients. Moreover, we also test this classification with a set of handcrafted texture features, classified using a Fully Connected Network (FCN), and a hybrid model combining both approaches.

\section{Related work}

Texture features such as intensity histogram statistics, the Gray-Level Co-occurrence Matrix (GLCM), the Run Length Matrix (RLM), Local Binary Patterns, and Wavelet-based features have been used in a few studies \cite{herlidou1999comparison, fan2014characteristics, duda2015mri, zhang2016optimal, eresen2019texture} towards the development of CAD methods for muscle diseases using MRI, particularly the Duchenne Muscle Dystrophy (DMD) and the Golden Retriever Muscle Dystrophy (GRMD), which is considered to be highly similar to DMD \cite{de2015application}.

More recently, Cai \textit{et al.}~\cite{cai2019texture} used different CNN architectures (ResNet and VGG) to automatically assess the progression of DMD and Congenital Muscular Dystrophy on Chemical Shift-Based Water-Fat Separation MRI. The authors also introduced an improved class activation mapping, which increased the mean accuracy of the best-performing CNN (ResNet-18) up to 91.7\%. Yang \textit{et al.}~\cite{yang2021deep} adopted a ResNet-50 model to diagnose dystrophinopathies on thigh muscle MRI, with an accuracy of 91\%. The authors also tested other CNN architectures, such as VGG-19, Inception-V3 and two DenseNet models. The lowest reported accuracy was 87\% (VGG-19).

Very little research has been published so far, regarding CAD methods for COL6-related myopathy. Bazaga \textit{et al.}~\cite{bazaga2019convolutional} proposed the first CAD method, which relies on a CNN to classify image patches, extracted from confocal microscopy images of fibroblast cultures.~A majority voting algorithm was applied to obtain a global decision on a given input image. The proposed method achieved an accuracy of 95\%. Recently, the authors of this paper assessed the effectiveness of several texture features in describing the level of muscle involvement on T1\emph{w} TSE MRI \cite{rodrigues2021isbi}. Feature selection was performed using SVM Recursive Feature Elimination (SVM-RFE) \cite{guyon2002gene}. The proposed method yielded accuracy values above 90\% for 3 of the 4 studied muscles.

\section{Methods}

\subsection{MRI data and patch extraction}

\begin{figure}[t!]
\centering
\includegraphics[width=0.95\linewidth]{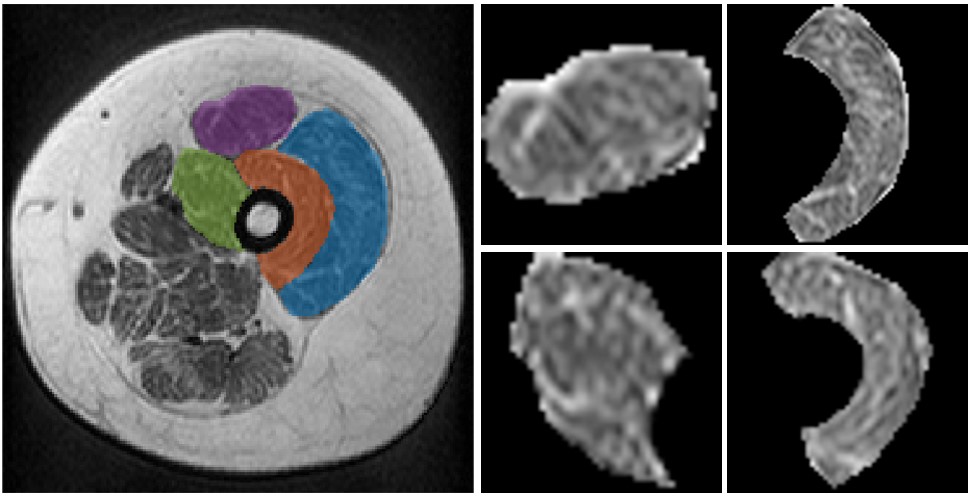}
\caption{Left: Segmentation of the \textit{Quadriceps femoris} muscles in an MR image of the thigh. Right: Image patches containing the segmented muscles.} \label{fig:muscle_segm}
\end{figure}

A group of 26 subjects, with ages ranging from 5 to 36 years old were examined using a 1.5T MRI scanner (Philips Medical Systems, Eindhoven, The Netherlands), at the Medical Imaging Department of the Raymond-Poincaré University Hospital (Garches, France).~UCMD or Bethlem Myopathy was diagnosed on 17 subjects, while the remaining 9 showed no perceivable signs of myopathy in the MR scans.

In this research, we selected slices from whole body muscle MRI scans performed in the same center, with the same magnet system and technical protocol~\cite{quijano2012whole}. We used axial T1\emph{w} TSE MR images of both thighs, with TR/TE = 631/16ms and slice thickness = 6mm. A different amount of slices were selected from each subject (6 to 12), to maximize the cross-sectional area of target muscles.

The \emph{Quadriceps femoris} muscles, i.e., \textit{Vastus lateralis}, \textit{Vastus medialis}, \textit{Vastus intermedius} and \textit{Rectus femoris}, were manually segmented in all selected images (Fig. \ref{fig:muscle_segm}, left). The data was annotated at the muscle level by experts, according to a 4-level scale proposed in \cite{mercuri2005muscle, lamminen1990magnetic} (healthy / no symptoms, mild, moderate, and severe). In this work, the scale was modified into a 3-level scale, by joining moderate and severe cases.

The selected T1\emph{w} TSE MRI dataset consisted of 196 \mbox{images}, which included both thighs. Thus, each of the four target muscles appeared in the dataset 392 times, yielding a total of 1568 regions of interest (ROI). Image patches containing the segmented muscles were obtained by cropping the MR image within the square segmentation bounding boxes (Fig. \ref{fig:muscle_segm}, right). All pixels outside the segmented muscles were set to 0. The size of the extracted patches ranged from 11x11 to 105x105. However, the vast majority were above 30x30. All patches were resized to 96x96, which was chosen as input size of the \mbox{\textit{ConvNet}} model described in Section \ref{sec:cnn}. This input size was chosen to promote a balance between avoiding texture distortions from resizing, as much as possible, and allowing a reasonable network depth for feature extraction.
 
\begin{figure*}[t!]
\centering
\includegraphics[trim={0 0 0.4cm 0}, clip, width=1\linewidth]{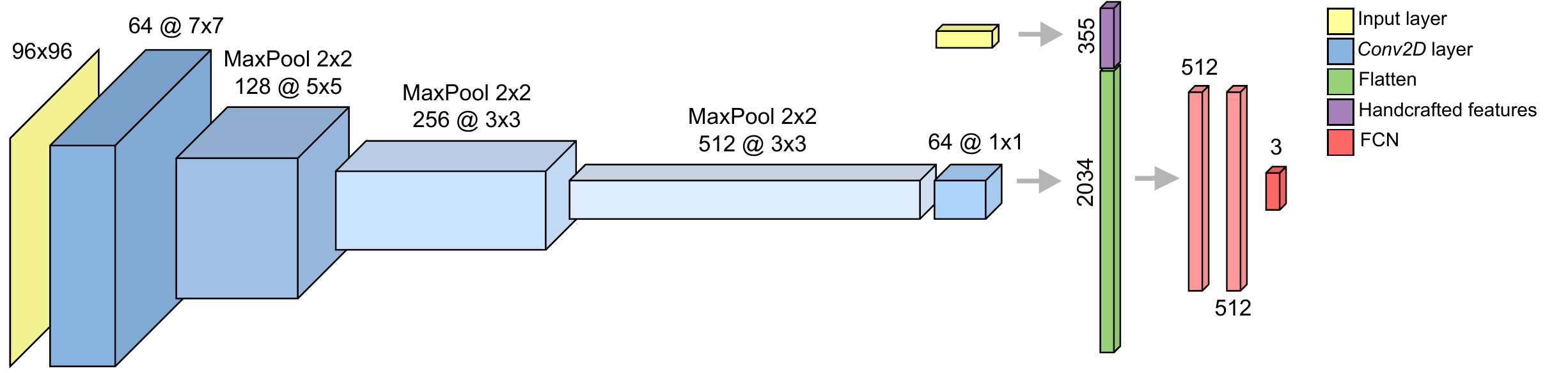}
\caption{Architecture of the hybrid classification model. In the \textit{ConvNet}-only model, the architecture is similar, except for the concatenation of the handcrafted texture features (top-right).} 
\label{fig:cnn_model}
\end{figure*} 

\begin{table*}
    \centering
    \caption{Performance results of the tested models. Precision, Recall, and the F-score are reported with respect to each class, whereas Accuracy refers to the overall classification accuracy.}
    \small
		\begin{tabular}{ccccccccccc}
		\toprule
		\multirow{2}{*}{\textbf{\normalsize Model}}					&	\multicolumn{3}{c}{\textbf{Healthy}} 			& \multicolumn{3}{c}{\textbf{Mild}} 			& \multicolumn{3}{c}{\textbf{Moderate/Severe}}  	& \multirow{2}{*}{\textbf{\normalsize Accuracy}} \\
		\cmidrule(lr){2-4} \cmidrule(lr){5-7} \cmidrule(lr){8-10}\\[-10pt]
		&	Precision &	Recall & 	F-score &  Precision &	Recall &  	F-score &	Precision &	Recall &  	F-score & 				   \\
	\midrule	
    \textbf{\textit{ConvNet}}   				&	0.96 &   	0.95 &   	0.96	&	0.73 &   	0.80 &   	0.76	&	0.95 &   	0.93 &   	0.94	&		91.3\% \\
    \textbf{Texture features}  		& 	0.98 &   	0.98 &   	0.98	&	0.73 &   	0.75 &  	0.74	&	0.93 &   	0.92 &   	0.92	&		90.8\% \\
	\textbf{Hybrid model}		&	0.98 &   	0.99  &  	0.99	&	0.82 &   	0.82	&	0.82  &	0.95 &   	0.94 &  	0.95	&		93.8\% \\
	\bottomrule
         
        \end{tabular}
        \label{tab:perf}
\end{table*}
 
\subsection{CNN model architecture}\label{sec:cnn}

For the classification of the disease severity in each segmented muscle, we implemented a \textit{ConvNet} encoder (Fig. \ref{fig:cnn_model}), consisting of sequential 2D convolutional layers, with an increasing number of filters (64, 128, 256, and 512), and progressively smaller spatial kernels (7x7, 5x5, 3x3, and 3x3). The output of the first three convolutional layers is downsampled using 2x2 max pooling. To further reduce the dimensionality of the encoded features, a 1x1 convolutional layer was added at the end. All convolutional layers used 'same' \textit{padding}, so that the height and width of their output matched those of the input. The classification part consisted of a FCN, with two layers of 512 nodes, both with 20\% \textit{dropout} to prevent \textit{overfitting}, and a 3-node output layer. The ReLU activation function was used in every layer, with the exception of the output layer, which used a \textit{softmax} activation.

\subsection{Handcrafted texture features}

As a follow-up to the study presented in \cite{rodrigues2021isbi}, we tested the classification of 355 ROI-based handcrafted texture features, obtained from the original images (i.e., without resizing), \mbox{using} an FCN classifier. These include statistics from the intensity and gradient histograms \cite{szczypinski2009mazda}, GLCM and RLM-based features \cite{eresen2019texture}, features from the autoregressive (AR) model \cite{szczypinski2009mazda}, Wavelet-based energy \cite{zhang2016optimal}, and statistics from the Gabor response magnitude images \cite{zhang2019adaptive}.

These texture features were computed using the \textit{MaZda} software \cite{szczypinski2009mazda}, with the exception of Gabor-based features, which were extracted using MATLAB R2020b. The FCN classifier included two densely connected layers of 256 nodes, using a ReLU activation, and a similar 3-node output layer with a \textit{softmax} activation.

\subsection{Hybrid model}

Finally, a hybrid classifier was also tested, which combines the handcrafted texture features and the flattened output of the \textit{ConvNet} into a single tensor. In this case, the FCN \mbox{architecture} was the same as in the \textit{ConvNet} experiment. The weights of the convolutional layers, which had been trained beforehand in each fold, were stored, and loaded for the \mbox{corresponding} test subset. The \textit{ConvNet} features in the \mbox{hybrid} model were computed without further training, whereas the parameters of the FCN classifier were fully trained in this experiment. Fig. \ref{fig:cnn_model} shows the full architecture diagram of the hybrid classification model.

\subsection{Model training and evaluation}

A leave-one-out cross-validation was set at the subject level for model training and evaluation, in order to maximize the number of patches in the training subset. 
Given the label imbalance within the dataset, class weights were computed in each fold, using the \textit{compute\_class\_weights} function of the \textit{scikit-learn Python} library. The weight for each label ($w_{l}$) was defined as $w_{l}= \nicefrac{N}{(L \times n_{l})}$, where $N$ and $L$ are the total number of images and labels, respectively, and $n_{l}$ is the number of images belonging to label $l$. The weights were passed as an argument to the fitting algorithm model, thus assigning a higher importance to patches of the least represented classes.

\begin{figure*}
     \centering
     \begin{subfigure}[b]{0.27\textwidth}
         \centering
         \includegraphics[width=\textwidth]{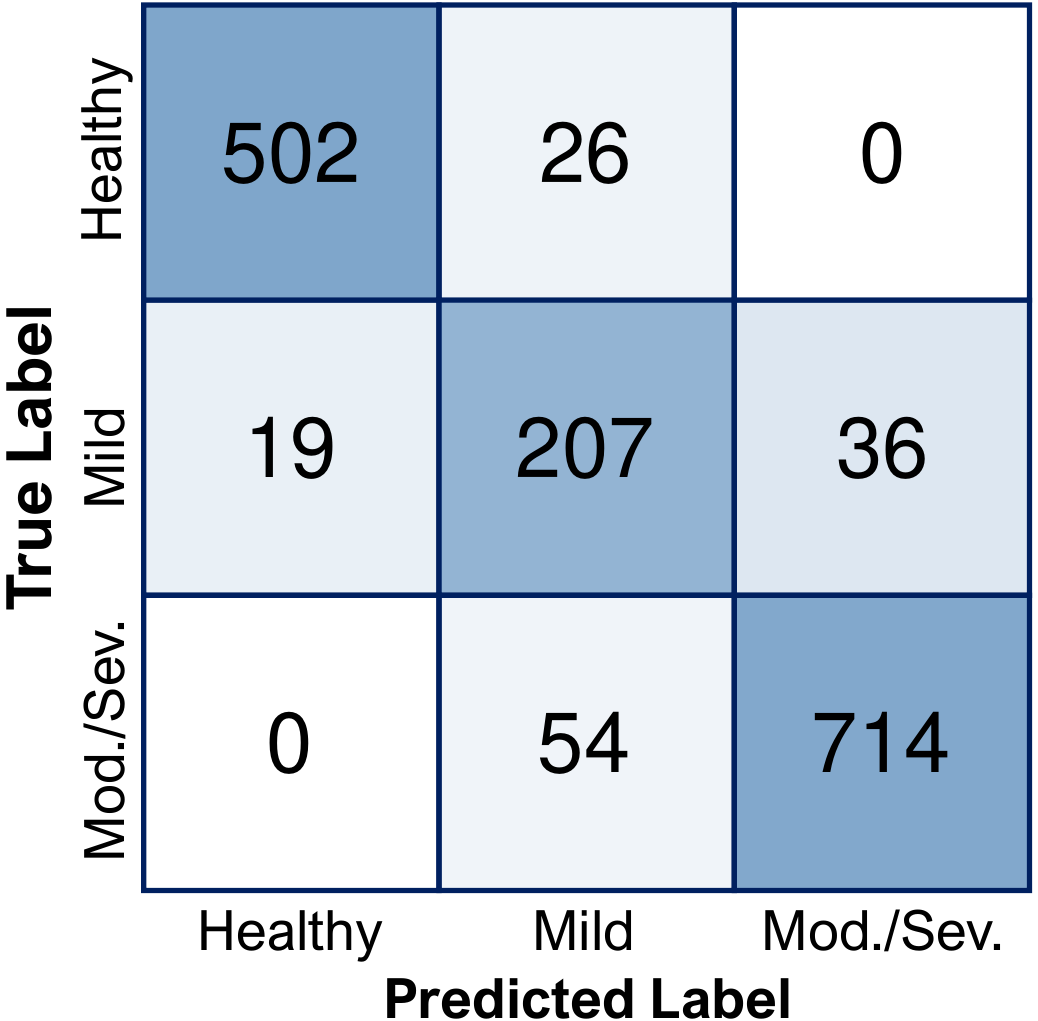}
         \caption{\textit{ConvNet}}
         \label{fig:cm_convnet}
     \end{subfigure}
     \hspace{0.6cm}
     \begin{subfigure}[b]{0.27\textwidth}
         \centering
         \includegraphics[width=\textwidth]{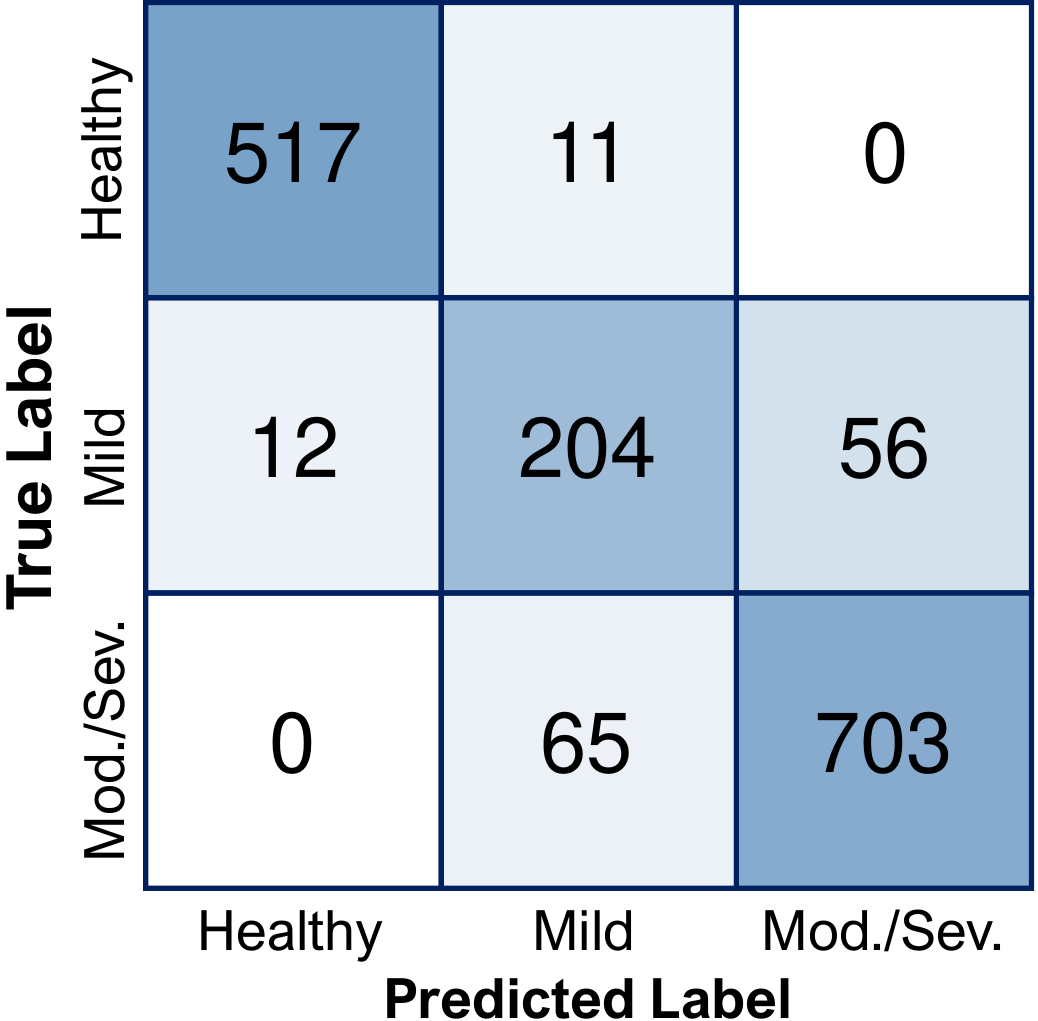}
         \caption{Texture features}
         \label{fig:cm_feats}
     \end{subfigure}
     \hspace{0.6cm}
     \begin{subfigure}[b]{0.27\textwidth}
         \centering
         \includegraphics[width=\textwidth]{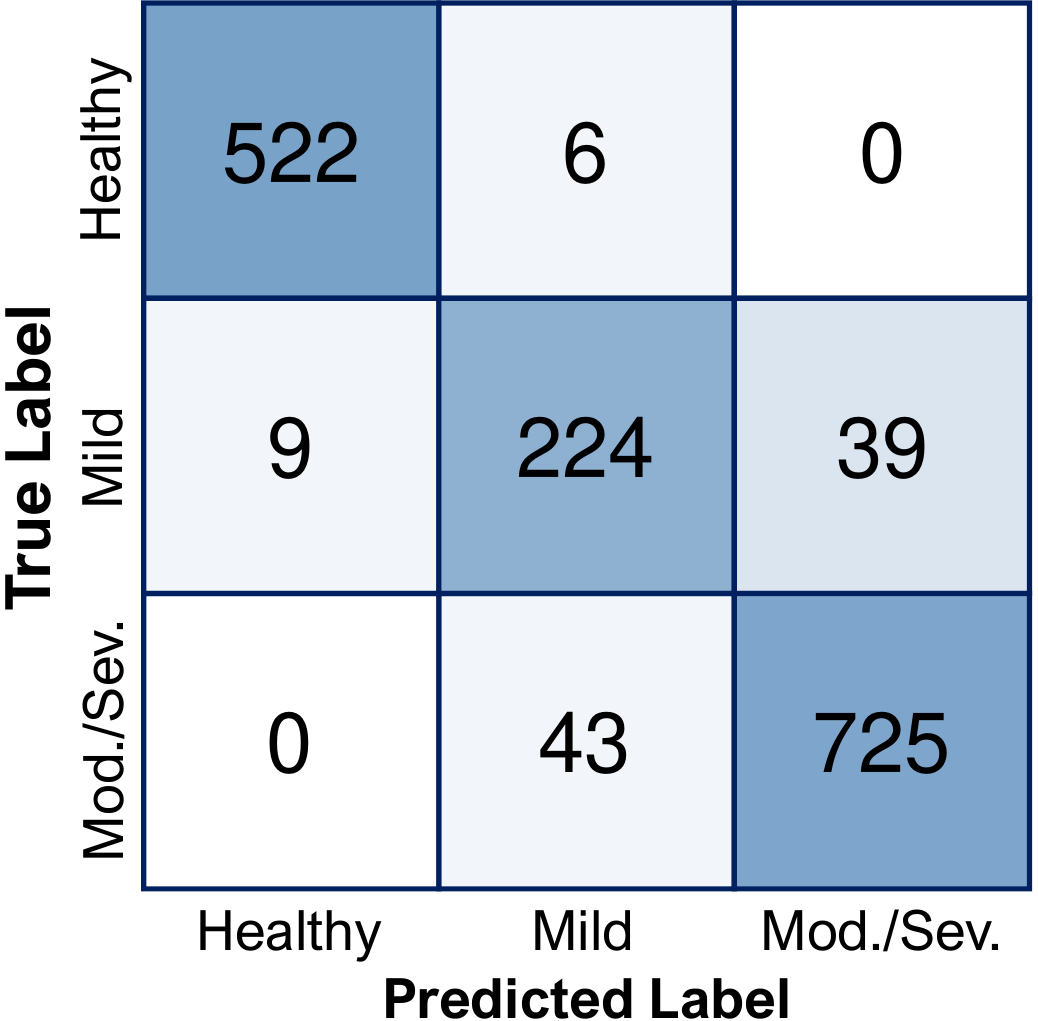}
         \caption{Hybrid model}
         \label{fig:cm_hybrid}
     \end{subfigure}
        \caption{Cross-validation confusion matrices.}
        \label{fig:conf_matrix}
\end{figure*}

\begin{table*}
    \centering
    \caption{Overall accuracy and average F-score (across all labels) considering each target muscle independently.}
    \small
		\begin{tabular}{ccccccccc}
		\toprule
		\multirow{2}{*}{\textbf{\small Model}}					&	\multicolumn{2}{c}{\textbf{\textit{Vastus lateralis}}} 			& \multicolumn{2}{c}{\textbf{\textit{Vastus intermedius}}} 			& \multicolumn{2}{c}{\textbf{\textit{Vastus medialis}}}  	& \multicolumn{2}{c}{\textbf{\textit{Rectus femoris}}}  \\
		\cmidrule(lr){2-3} \cmidrule(lr){4-5} \cmidrule(lr){6-7} \cmidrule(lr){8-9}\\[-10pt]
		&	Accuracy &	F-score & Accuracy & F-score & Accuracy & F-score &	Accuracy &  	F-score 				   \\
	\midrule	
    \textbf{\textit{ConvNet}}   		&	94.6\% &   	0.92 &   	93.4\%	&	0.89 &   	83.9\% &   	0.80	&	93.6\% &   	0.93  \\
    \textbf{Texture features}  		&	89.3\% &   	0.84 &   	96.4\%	&	0.94 &   	85.0\% &   	0.82	&	94.1\% &   	0.93  \\
	\textbf{Hybrid model}		&	94.4\% &   	0.91 &   	97.0\%	&	0.94 &   	87.5\% &   	0.85	&	96.4\% &   	0.96  \\
	\bottomrule
         
        \end{tabular}
        \label{tab:musc}
\end{table*}

To address the issue of poor network generalization that might arise from a relatively small dataset, we implemented a data augmentation scheme. 
At each iteration the images were transformed through the random application of horizontal flipping, horizontal translation within a range of [-10\%, 10\%] of the image width, rotation within a range of [-0.2, 0.2] $\times 2\pi$, and contrast adjustment with a contrast factor ($c_f$) within a range of [0.8, 1.2]. 
The pixel values $x$ are adjusted according to $c_f(x - \overline{x}) + \overline{x}$ ), where $\overline{x}$ represents the mean of the pixel values. The grayscale input images were also \mbox{normalized} to the [0, 1] range.

The proposed models were implemented using the Keras API, with a Tensorflow backend, and optimized using the Adam algorithm for 100 epochs, with a batch size of 25, using the categorical cross-entropy loss function. The learning rate was initially set at 10\textsuperscript{-3}, with a step-wise decay of 0.1 at every 20 epochs. We also implemented L2 weight regularization  with $\alpha$ = 0.01, to further help reducing the probability of \textit{overfitting}.

\section{Results and discussion}

Table \ref{tab:perf} shows performance measures of the three tested models (i.e., precision, recall, and F-score) considering each severity grade, as well as the global accuracy. The hybrid model achieves better results, with a higher global classification accuracy of 93.8\%, mainly due to the better classification of mild cases. 
A comparison between the \textit{ConvNet} and the texture features classifier shows that the first performed better with the mild and moderate/severe cases (Fig. \ref{fig:cm_convnet}), while the second improved the correct identification of healthy cases (Fig. \ref{fig:cm_feats}). All models performed well in separating healthy from affected cases, even between healthy and mild, which shows good potential for the early detection of myopathy without biopsy. From the confusion matrices in Fig. \ref{fig:conf_matrix}, we may conclude that misclassifications occurred only between adjacent classes, i.e., healthy/mild or mild/moderate-severe. This suggests the potential of using the proposed approaches in follow-up studies, e.g., in monitoring treatment outcome with reduced invasiveness.

Considering only results for each muscle, which are summarized in Table \ref{tab:musc}, the \textit{Vastus medialis} led to the worst \mbox{overall} results, and particularly in mild cases with 32 misclassifications (\textit{ConvNet}), 31 (texture features), and 26 (hybrid model), out of 80 patches. 
Most errors were mild samples classified as moderate/severe, but a few cases of mild classified as healthy also occurred (7 with \textit{ConvNet} and 1 with the other two \mbox{methods}. These results are in line with those \mbox{obtained} in \cite{rodrigues2021isbi}, where the \textit{Vastus medialis} also led to the the worst results (recall = 0.77 / precision = 0.72) considering the three classes. Fewer misclassifications of mild cases occurred in other muscles, with either model (maximum of 15 with the texture features for the \textit{Rectus femoris}). However, the smaller number of samples, when compared with the healthy and moderate/severe classes, accentuates the weight of these errors in the performance measures.

This study differs from \cite{rodrigues2021isbi}, as ROI patches from all muscles were combined in model training, with the intention of working towards a more generalized and automated solution. Nonetheless, as is common with deep learning-based solutions, the lack of training data is very likely to be a relevant constraint to the improvement of these results. 

\section{Conclusions and future work}

The proposed models have shown potential to assist radiologists in performing non-invasive diagnosis and follow-up of patients with COL6-related myopathies, as well as other NMD. Because these diseases are fortunately rare, it is even more important to find solutions that provide accurate \mbox{classifications} with reduced amounts of data.

In future research efforts, it will be essential to test this approach with an increased number of different muscles, towards enabling fully automated whole body myopathy screenings, combined with a muscle segmentation method.

\bibliographystyle{IEEEbib}
\bibliography{refsICIP2022}

\end{document}